\documentclass[conference]{IEEEtran}
\IEEEoverridecommandlockouts

\usepackage{blindtext, graphicx}
\usepackage{listings}
\usepackage{algorithm}
\usepackage{algorithmicx}
\usepackage{amsmath,amssymb,amsfonts} %
\usepackage{textcomp} %
\usepackage{subfig} %
\usepackage{multirow}
\usepackage{cite}
\usepackage{url}
\usepackage{amsthm}
\usepackage{booktabs} 
\usepackage{framed}
\usepackage{color}
\usepackage{xcolor}
\usepackage[skip=1pt]{caption}
\usepackage{tabularx}
\usepackage{rotating}
\usepackage[bookmarks=false]{hyperref}
\usepackage{enumerate}
\usepackage{comment}
\usepackage{fancyhdr}
\usepackage{lipsum}
\usepackage{changepage}
\usepackage{enumitem}

\algdef{SE}[DOWHILE]{Do}{doWhile}{\algorithmicdo}[1]{\algorithmicwhile\ #1}

\usepackage{algorithm}
\usepackage{algorithmicx}
\usepackage[noend]{algpseudocode}

\algnewcommand{\IfInline}[1]{\State\algorithmicif\ \, #1\ \, \algorithmicthen \, }
\algnewcommand{\EndIfInline}{\unskip\ }

\usepackage{amsthm}
\usepackage{amssymb}
\usepackage{booktabs} 
\usepackage{framed}
\usepackage{url}
\usepackage{color}
\usepackage{xcolor}
\usepackage{subfig}
\usepackage[skip=1pt]{caption}
\usepackage{tabularx}
\usepackage{rotating}
\usepackage{url}
\usepackage{hyperref}
\usepackage{amsmath}
\usepackage{multirow}
\usepackage{enumerate}
\usepackage{comment}
\usepackage{fancyhdr}
\usepackage{lipsum}
\usepackage{changepage}
\usepackage{enumitem}
\usepackage{balance}

\algdef{SE}[DOWHILE]{Do}{doWhile}{\algorithmicdo}[1]{\algorithmicwhile\ #1}

\newcommand{\cE}{{\mathcal E}}

\newcommand{\cG}{{\mathcal G}}
\newcommand{\cH}{{\mathcal H}}

\newcommand{\cV}{{\mathcal V}}

\newcommand{\cS}{{\mathcal S}}

\newcounter{alg}
\def\BibTeX{{\rm B\kern-.05em{\sc i\kern-.025em b}\kern-.08em
    T\kern-.1667em\lower.7ex\hbox{E}\kern-.125emX}}

\ifCLASSINFOpdf
\else
\fi

\begin{document}
%
%
\title{Towards Distributed 2-Approximation Steiner Minimal Trees in Billion-edge Graphs
}

%
%

%
\author{
\IEEEauthorblockN{Tahsin Reza, Geoffrey Sanders and Roger Pearce}
\IEEEauthorblockA{Center for Applied Scientific Computing (CASC), Lawrence Livermore National Laboratory (LLNL) \\
Email: \{reza2, sanders29, rpearce\}@llnl.gov
}
}

\maketitle
\begin{abstract}
%
Given an edge-weighted graph and a set of known {\em seed} vertices of interest, a network scientist often desires to understand the graph relationships to {\em explain} connections between the seed vertices. If the size of the seed set is 2, shortest path calculations are an attractive computational kernel to explore the connections between the two vertices. When the seed set is 3 or larger (say up to 1,000s) {\em Steiner minimal tree} -- min-weight acyclic connected subgraph (of the input graph) that contains all the seed vertices -- is an attractive generalization of shortest weighted paths. In general, computing a Steiner minimal tree is NP-hard, but decades ago several polynomial-time algorithms were designed and proven to yield Steiner trees whose total weight is bounded within 2 times the minimal Steiner tree.

Despite its rich theoretical literature, works related to parallel Steiner 
minimal tree computation and their scalable implementations are rather scarce. In this paper, we present a parallel 2-approximation Steiner minimal tree algorithm (with theoretical guarantees) and its MPI-based distributed implementation. In place of distance computation between all pairs of seed 
vertices, an expensive phase in many approximation algorithms, the solution we employ, exploits {\em Voronoi cell} computation. Also, this approach has higher parallel efficiency than others that involve minimum spanning tree computation on the entire graph. Furthermore, our distributed design exploits asynchronous processing and a message prioritization scheme to accelerate 
convergence of distance computation, employs techniques to avoid inefficient distributed spanning tree computation on the entire graph, and harnesses a combination of vertex and edge centric processing to offer fast time-to-solution. We demonstrate scalability and performance of our solution using real-world graphs with up to 128 billion edges and 512 compute nodes (8K processes), show the ability to find Steiner trees with up to 10K seed vertices in under one minute, and present in-depth analyses that highlight 
the benefits of our design choices. Using four real-world graphs and three seed sets for each,  we compare our solution with the state-of-the-art exact Steiner minimal tree solver, SCIP-Jack, and two sequential algorithms with the same approximation bound as our algorithm. Our distributed solution comfortably outperforms these related works on graphs with 10s million edges and offers decent strong scaling -- up to 90\% efficient. We empirically show that, on average, the total distance (sum of edge weights) of the Steiner tree identified by our solution is 1.0527 times greater than the Steiner minimal tree (i.e., the optimal solution) -- well within the theoretical bound of less than equal to 2.

\end{abstract}



%
%
\section{Introduction}
\label{section:introduction}

Networks are often represented by a {\em distance-weighted} graph $\cG(\cV, \cE, d)$, with data entities represented by vertices $\cV$, 
their relationships represented by edges $\cE \subset \cV \times \cV$, and distances $d : \cE \rightarrow [1, \infty)$. In this work, smaller weights represent stronger relationships (or closer distances between the underlying data entities). For $\{u, v\} \in \cV$, the {\em weight} or {\em distance} of $(u, v) \in \cE$ is written $d(u, v)$, and for a knowledge network, is often a function of the metadata living on $u$, $v$, $(u, v)$ and the relationship type of $(u, v)$. The total weight or total distance of a set of edges associated with a subgraph $\cH$ (e.g., a cluster, a tree or a path) is the sum of the distances $D(\cH) = \sum_{(u, v) \in \cH} d(u, v)$. We call the user's entities of interest, the {\em seed} vertices, $\cS \subset \cV$, and various applications need different magnitude of seed set sizes. The goal is to compute a relatively small subgraph $\cG_\cS$ that connects all vertices in $\cS$, preferring edges that have low distance over those that are larger -- known in the literature as the {\em Steiner minimal tree} problem~\cite{Karp1972}.

\begin{figure}[!t]
\centering
\includegraphics[width=\linewidth]{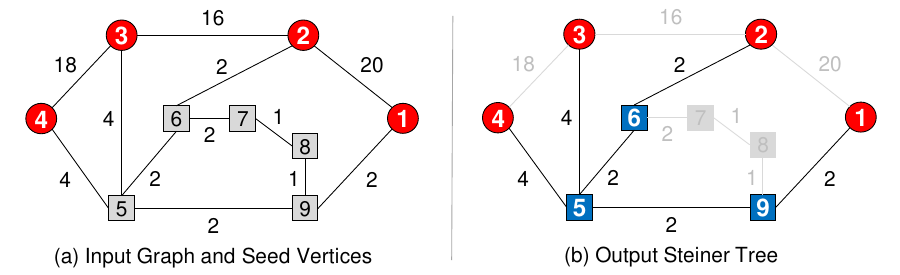} 
\setlength{\belowcaptionskip}{-12pt}
\captionsetup{font=footnotesize}
\caption{(a) A data graph $\cG$, and given seed vertices $\cS$, have red fill. (b) A Steiner tree $\cG_\cS$ of $\cG$. Steiner vertices $\cS' \in \cV \setminus \cS$ have blue fill. Vertices and edges not in the Steiner tree $\cG_\cS$ have light grey outline and fill, respectively.}
\label{fig:asmt_example_0011}
\end{figure}

Given a set of seed vertices $\cS$ (also called {\em terminal} vertices in the literature), a Steiner tree $\cG_\cS(\cV_\cS, \cE_\cS, d_\cS)$ of $\cG(\cV,\cE,d)$ is called a Steiner minimal tree, if its total distance $D(\cG_\cS)=\sum_{e\in\cE_\cS}d_\cS(e)$ is minimal among all Steiner trees for $\cG$ and $\cS \subset \cV$. In the literature, the vertex set $\cS' \subset \cV \setminus \cS$ is commonly refereed as {\em Steiner vertices}, although are not required to be in a Steiner tree, 
may be used to achieve a small total distance~\cite{Gilbert.1968.doi:10.1137/0116001,Hakimi1971SteinersPI,Karp1972}. Fig.~\ref{fig:asmt_example_0011} shows an example. Gilbert and Pollack~\cite{Gilbert.1968.doi:10.1137/0116001} are often credited for formalising the Steiner minimal tree problem which has direct applications to VLSI design~\cite{Ihler.1999.10.1016/S0166-218X(98)00090-0, Caldwell.1999.784119}, communication network optimization~\cite{Sun.2019.10.1109/TNET.2018.2888864, Gong.MobiHoc.2015.10.1145/2746285.2746296}, computational and systems biology~\cite{Sun.2017.BMC.s12859-017-1958-4, Lu.2002.10.1007/3-540-45655-4_13}, and knowledge discovery and management~\cite{Lee.2012.10.1016/j.cie.2011.11.013, Li.2016.10.1145/2882903.2915217}. The problem of finding a Steiner minimal tree of a graph and arbitrary seed vertices is known to be NP-hard (the decision variant is NP-complete)~\cite{Karp1972, Garey.1976.10.1145/800113.803626}. Therefore, polynomial time algorithms for finding a Steiner tree $\cG_\cS$ with a total distance $D(\cG_\cS)$ close to the total distance of a Steiner minimal tree $D_{min}(\cG)$ are sought due to their practical relevance.

Often in large knowledge networks, property graphs, or other relational datasets, a user seeks to understand the relationships between two or more entities of interest. 
Without knowing much about the graph and entities of interest a priori, it is difficult to know precisely what analytics and parameters gives users exactly what they want, and an interactive framework is highly desired for exploring data relationships. Today's knowledge networks are also massive, so this framework needs to be scalable and efficient enough to provide palatable interactivity. We aim at a framework that is scalable in a parallel, distributed computing 
environment, where the topology and metadata of massive relational datasets can be stored concurrently in memory. 

Typically, a user will interact with such computation in various ways, exploring the relationships, as 
several factors dramatically change the nature of the output size (for example, the number of graph hops or shortest path distances between pairs of vertices and the presence of multiple paths that are tied in hop length and distance). This includes the user adding or removing classes of edges and/or vertices and adjusting edge distance functions based on investigating the output. Such interaction warrants computations that 
can be made as fast as possible and strongly scale, so more computing resources can be employed to
speed up the calculations when needed.

When $|\cS| = 2$, sets of edges that exist in shortest weighted paths and near-shortest weighted paths (low total distance paths) provide an attractive framework for understanding the relationships between the seeds. Computing edge and vertex sets with such paths is fairly easy, even for massive graphs using distributed computing,  
and simple variants allow the user to remove edges (rank vertices and edges in various ways and remove the low-ranking subgraph members) or add edges of near-shortest paths (i.e., augmenting paths) to build up a subgraph. When $|\cS| > 2$, low-weight Steiner trees 
provide an attractive framework and minimally-weighted Steiner trees are a direct generalization of shortest weighted paths.

Since Steiner minimal tree is an NP-hard problem, there have been continuing interests in practical polynomial-time solutions with tight approximation bound. Takahashi et al.~\cite{takahashi1980approximate} presents an algorithm with the approximation bound $D(\cG_\cS)/D_{min}(\cG)\leq2(1-1/|\cS|)$. The algorithm by Kou et al.~\cite{kou.1981.10.1007/BF00288961} (known as the KMB algorithm) improves the bound to $D(\cG_\cS)/D_{min}(\cG)\leq2(1-1/l)$ where $l$ is the minimum number of leaves in any Steiner minimal tree for $\cG$ and $\cS$. A corpus of algorithms capitalizes over the KMB algorithm and offer improved sequential runtime-complexity while preserving the 2-approximation bound: algorithms by Wu et al.~\cite{Wu1986FA} (known as the WWW algorithm), Widmayer~\cite{Widmayer.1987.10.1007/3-540-17218-1_46} and Mehlhorn~\cite{MEHLHORN1988125} are some the most well known 2-approximation solutions.

Despite its rich theoretical literature, work related to parallel Steiner minimal tree computation and their scalable implementations is rather scarce. In this paper, we present a parallel 2-approximation Steiner minimal tree algorithm and its MPI-based distributed implementation. Our solution is based on the idea of computing {\em Voronoi cells} similar to~\cite{MEHLHORN1988125}: Mehlhorn replaces the most computationally expensive task in the KMB algorithm, i.e., computing all-pair-shortest-paths (APSP) between the vertices in $\cS$, by Voronoi cell computation of each $s\in\cS$. Our solution approach stems from the observations that, in practice, the Voronoi cell approach is less expensive then APSP computation (see Table~\ref{table:evaluation_compare_apsp_voronoi_cell}), and has higher parallel efficiency than minimum spanning tree (MST) computation on the entire graph~\cite{Bader.IPDPS.2004.1302953} (as in WWW and Widmayer algorithms). To accommodate large-scale graphs, e.g., 100s billion edges, we adopt a scale-out design based on graph partitioning. Our distributed design embraces a combination of vertex and edge centric processing, asynchronous processing, and a message prioritization scheme to offer fast time-to-solution. Below we summarize the key contributions in this paper.

\begin{itemize}[leftmargin=16pt]

\item[\textit{(i)}] We present a parallel algorithm, based on Voronoi cell identification for distance computation~\cite{MEHLHORN1988125}, for 2-approximation Steiner minimal tree computation that offers guarantees on the approximation bound (\S\ref{section:preliminaries} and \S\ref{section:solution_approach}). 

\item[\textit{(ii)}] We present the design of a proof-of-concept distributed implementation of our parallel Steiner tree algorithm. The solution, \textit{(a)} is designed to accommodate large graphs through data partitioning; \textit{(b)} performs both vertex and edge centric processing to achieve fast time-to-solution; \textit{(c)} embraces asynchronous processing and message prioritization to accelerate convergence of distance computation and achieves message efficiency in the process; \textit{(d)} employs techniques to avoid inefficient distributed minimum spanning tree computation on the entire graph (as some of the key sequential algorithms do~\cite{Wu1986FA, Widmayer.1987.10.1007/3-540-17218-1_46}); \textit{(e)} extends \mbox{HavoqGT}~\cite{Pearce:2014:FPT:2683593.2683654}, an MPI-based vertex-centric graph framework, for algorithm implementation (\S\ref{section:system_description}). 

\item[\textit{(iii)}] We demonstrate scalability and performance using eight real-world graphs with up to 128 billion edges and up to 512 compute nodes (8K processes) -- up to 90\% efficient strong scaling, t.t.b.o.o.k., the largest scale to date for the Steiner tree problem; show the ability to find Steiner trees with up to 10K seed vertices in under one minute; evaluate the effectiveness of our design choices and optimizations (\S\ref{section:evaluation}). Using 12 data instances, we compare our solution with the state-of-the-art exact Steiner tree solver, SCIP-Jack~\cite{Rehfeldt.2021.10.1007/978-3-030-55240-4_10}, and two 2-approximation sequential algorithms, WWW~\cite{Wu1986FA} and Mehlhorn~\cite{MEHLHORN1988125}. 
Our distributed solution comfortably outperforms these related works on graphs with 10s million edges. On average, the total distance $D(\cG_\cS)$ of the Steiner tree identified by our solution is 1.0527 times greater than the Steiner minimal tree $D_{min}(\cG)$, which is well within the theoretical bound of $\leq 2(1 - 1/l)$, and the approximation error is 5.3\% (\S\ref{section:evaluation_compare_related_work}).  
\end{itemize}

\begin{table}[!h]
\footnotesize
\renewcommand{\arraystretch}{1.2}
\captionsetup{font=footnotesize}
\caption{Runtime comparison of all-pair-shortest-path (APSP) and Voronoi cell (VC) computation using two graphs and three seed sets. All experiments use a single thread. Dataset details are in \S\ref{section:evaluation}, Table~\ref{table:evaluation_datasets}.}
\label{table:evaluation_compare_apsp_voronoi_cell}
\setlength\tabcolsep{3.0pt}
\centering
\begin{tabular}{l|rr|rr|rr}
\hline
$|S|$  & \multicolumn{2}{c|}{10} & \multicolumn{2}{c|}{100} & \multicolumn{2}{c}{1000} \\
\hline
& APSP & VC & APSP & VC & APSP & VC \\
\hline
LVJ & 49.7s & 30.0s & 539.2s & 35.1s & 5,813.3s & 104.5s \\
PTN & 26.7s & 12.9s & 270.3s & 26.6s & 2,767.4s & 85.5s \\
\hline
\end{tabular}
\end{table}

%
\section{Preliminaries}
\label{section:preliminaries}

Table~\ref{table:preliminaries_notation} lists the symbolic notation used in this paper, following the convention in~\cite{kou.1981.10.1007/BF00288961,Wu1986FA,Widmayer.1987.10.1007/3-540-17218-1_46,MEHLHORN1988125}. A {\em path} is a non-repeating sequence of vertices in $\cV$. A vertex in a path has at least one, at most two adjacent vertices. $d_1(v_i, v_j)$ is equal to the distance of a shortest path from $v_i$ to $v_j$ in $\cG$. For every seed vertex $s\in\cS$, {\em Voronoi cell} $N(s)$ is the set of vertices in $\cV$ that are at a shorter distance (an edge or a path) to $s$ than to any other vertex in $\cS$. $N(s) \cap N(t) = \emptyset$ for any $s,t\in\cS$ and $s \neq t$, with $v\in N(s) \Rightarrow d_1(v,s) \leq d_1(v,t)$. An edge $(u, v) \in \cE$ is a {\em cross-cell edge} if $u\in N(s)$, $v\in N(t)$, and $s, t \in \cS$, $s \neq t$. A cross-cell edge bridges two Voronoi cells.  

Kou et al. were the first to propose an algorithm (Alg.~\ref{alg:kmb}) for finding a Steiner tree $\cG''$ with \mbox{$D(\cG'')/D_{min}(\cG)\leq2(1-1/l)$}, where $l$ is the minimum number of leaves in any Steiner minimal tree for $\cG$ and $\cS$. The most expensive step in the KMB algorithm (Alg.~\ref{alg:kmb}) is all-pair-shortest-path computation among the seed vertices to form $\cG_1$, the dense graph of minimal distance between all seeds. Mehlhorn~\cite{MEHLHORN1988125} proposes to replace Step 1. in Alg.~\ref{alg:kmb} by a cheaper alternative which is based on Voronoi cell computation of every $s\in\cS$. Mehlhorn improves sequential complexity of KMB from $O(|\cS||\cV|^2)$ to $O(|\cV|log|\cV|+|\cE|)$ with an algorithm that is more amenable to parallel, distributed computing.

Mehlhorn observes that it is possible to construct a distance graph $\cG'_1$ which is a subgraph of $\cG_1$ in Alg.~\ref{alg:kmb}, where $\cG'_1(\cS,\cE'_1,d'_1)$ is defined by:

$\cE'_1 = \{(s,t); s,t \in \cS$ and there is an edge $(u,v) \in \cE$ with 
\indent $u \in N(s)$, $v \in N(t)\}$ and

$d'_1(s,t) = min(d_1(s,u) + d(u,v) + d_1(v,t)); \, (u,v) \in \cE,$ \indent \mbox{$\, u \in N(s), \, v \in N(t)$}.

\noindent Note: in general, $d'_1$ is not the restriction of $d_1$ to the set $\cE'_1$. 

Mehlhorn proves that existence of an MST $\cG_2$ of $\cG_1$ that is also a subgraph of $\cG'_1$, that the associated distance functions $d_1$ and $d_1'$ agree on edges of $\cG_2$, and further, that every MST of $\cG'_1$ is also a MST of $\cG_1$. This fact allows approximation bound of the KMB algorithm to apply to Mehlhorn's algorithm.

\begin{table}[!t]
\footnotesize
\renewcommand{\arraystretch}{1.2}
\captionsetup{font=footnotesize}
\caption{Symbolic notation used.}
\label{table:preliminaries_notation}
\setlength\tabcolsep{3.0pt}
\centering
\begin{tabular}{l|l}
\hline
Object(s) & Notation \\ 
\hline
background graph, vertices, edges & $\cG (\cV, \cE, d)$ \\
background graph vertices & $\cV := \{v_0, v_1, ..., v_{n-1}\}$\\
background graph edges  & $(v_i, v_j) \in \cE$\\
set of vertices adjacent to $v_i$ in $\cG$ & $adj(v_i)$\\
predecessor, successor of $v_i$ & $pred(v_i)$, $scsr(v_i)$ \\
$d$ is a distance function which  & \\ 
\hspace{6pt}maps $\cE$ in to a set of non-zero, & $d(v_i,v_j) \in \mathbb{Z}^{+} \setminus 0$ \\ 
\hspace{6pt}non-negative integers & \\
\hline
Steiner tree, vertices, edges, & $\cG_\cS (\cV_\cS, \cE_\cS, d_\cS)$, $\cV_\cS\subset\cV$, $\cE_\cS\subset\cE$ \\
seed vertices, Steiner vertices & $\cS\subset\cV_\cS$, $\cS' \subset \cV \setminus \cS$ \\
$d_\cS$ is the distance function for $\cG_\cS$ & $d_\cS(v_i,v_j) \in \mathbb{Z}^{+} \setminus 0$, $(v_i,v_j)\in\cE_\cS$\\
\hline
Voronoi cell of vertex $s\in\cS$ & $N(s)$, $\cV = \bigcup_{s\in\cS}N(s)$ \\
source vertex of $v_i$ & $src(v_i) = s$ if $v_i \in N(s)$\\
\hline
total distance of $\cG_\cS$ & $D(\cG_\cS) = \sum_{e\in\cE_\cS} d_\cS(e)$ \\
distance of Steiner minimal tree & $D_{min}(\cG)$ \\
approximation ratio & 
$D(\cG_\cS)/D_{min}(\cG)$ \\
\hline
\end{tabular}
\end{table}

\begin{algorithm}[!h]
\small
\caption{KMB Algorithm~\cite{kou.1981.10.1007/BF00288961}}
\label{alg:kmb}
\hspace*{\algorithmicindent}\textbf{Input:} edge-weighted graph $\cG$, seed vertices $\cS$ \\
\hspace*{\algorithmicindent}\textbf{Output:} Steiner tree $\cG_5$
\begin{algorithmic}[1]
\State Construct the complete distance graph $\cG_1(\cV_1,\cE_1,d_1)$ where $\cV_1=\cS$ and, for every $(v_i,v_j)\in\cE_1$, $d_1(v_i,v_j)$ is equal to the distance of a shortest path from $v_i$ to $v_j$ in $\cG$.
\State Find MST $\cG_2$ of $\cG_1$. 
\State Construct a subgraph $\cG_3$ of $\cG$ by replacing each edge in $\cG_2$ by (one of) the corresponding shortest path(s) in $\cG$.
\State Find MST $\cG_4$ of $\cG_3$ .
\State Construct a Steiner tree $\cG_5$ from $\cG_4$ by deleting edges in $\cG_4$, if necessary, so that no leaves in $\cG_5$ are Steiner vertices. 
\end{algorithmic}
\end{algorithm}

\section{Parallel Steiner Tree Algorithm}
\label{section:solution_approach}

In this section, we introduce our parallel Steiner tree algorithm, while in \S\ref{section:system_description}, we discuss the distributed implementation. The algorithm produces a Steiner tree $\cG_\cS$ with the approximation ratio $D(\cG_\cS)/D_{min}(\cG)\leq2(1-1/l)$, same as~\cite{kou.1981.10.1007/BF00288961, Wu1986FA, Widmayer.1987.10.1007/3-540-17218-1_46, MEHLHORN1988125}. Our solution is based on the idea of computing Voronoi cells similar to~\cite{MEHLHORN1988125}. The generalized minimum spanning tree computation approach~\cite{Wu1986FA, Widmayer.1987.10.1007/3-540-17218-1_46} may appear attractive because of its simplicity and greater work efficiency, however, we observe that it is possible to develop an algorithm with higher parallel efficiency for Voronoi cell computation compared to that of MST computation. Bader et al.~\cite{Bader.IPDPS.2004.1302953} and authors of the Galois~\cite{Dathathri.2019.PCT.8891625} project demonstrated MST computation suffers from rapid decrease in the available parallelism~\cite{galois.mst.2016}. 

\begin{figure*}[!t]
\centering
\includegraphics[width=\linewidth]{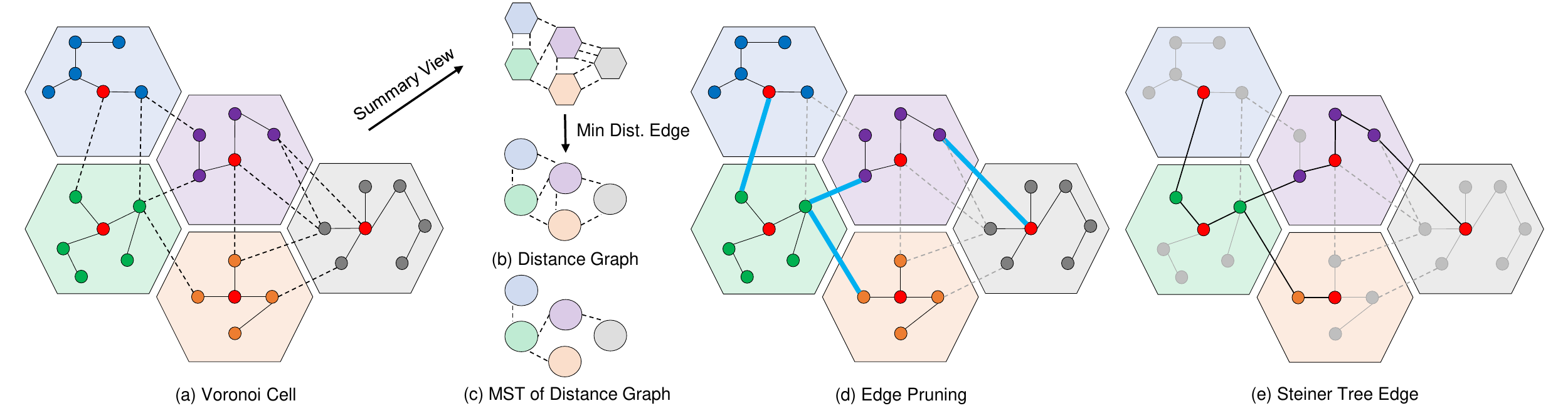} 
\setlength{\belowcaptionskip}{-12pt}
\captionsetup{font=footnotesize}
\caption{High-level illustration of output of key algorithm steps: (a) Each hexagon represents the Voronoi cell of a seed vertex (solid red fill); cross-cell edges are shown using broken lines. (b) The distance graph $\cG'_1$. (c) MST $\cG'_2$ of $\cG'_1$. (d) Post MST edge pruning: ``deleted" cross-cell edges have light grey outline; ``active" cross-cell edges (corresponding edges of MST $\cG'_2$ in $\cG$) are shown using blue outline. (e) The vertices and edges in the final Steiner tree.}
\label{fig:asmt_algorithm_steps_0011}
\end{figure*}

\begin{algorithm}[!b]
\small
\caption{Parallel Steiner Tree Algorithm}
\label{alg:parallel_steiner_tree}
\hspace*{\algorithmicindent}\textbf{Input:} edge-weighted graph $\cG$, seed vertices $\cS$ \\
\hspace*{\algorithmicindent}\textbf{Output:} Steiner tree $\cG_\cS$, total distance $D(\cG_\cS)$
\begin{algorithmic}[1]
\State In parallel, for every $s \in \cS$, compute Voronoi cell $N(s)$. Every $u \in \cV$, $u \in N(s)$,  maintains states $src(u)$, $pred(u)$, $d_1(s, u)$.

\State Construct the distance graph $\cG'_1(\cS,\cE'_1,d'_1)$: In parallel, for every $(s,t)\in\cE'_1$, compute \mbox{$d'_1(s,t) = min(d_1(s,u)+d(u,v)+d_1(v,t))$}, where $s, t \in \cS$, $s \neq t$ and $u\in N(s)$, $v\in N(t)$. 

\State Find MST $\cG'_2$ of $\cG'_1$ using a sequential algorithm.

\State In parallel, for every cross-cell edge $(u, v) \in \cE$,
mark $(u, v)$ as ``deleted", where $(s, t) \in \cG'_1$, $d'_1(s,t)=d_1(s,u)+d(u,v)+d_1(v,t)$ and $(s, t) \notin \cG'_2$.

\State In parallel, for every ``active'' crocess-cell edge $(u,v) \in \cE$, where $u \in N(s)$, $v \in N(t)$ and $s \neq t$, identify all edges in the shortest path from $u$ to $s$, and $v$ to $t$, by following respective predecessor vertices stored in $pred$.  

\State $\cG_\cS(\cV_\cS,\cE_\cS,d_\cS)$ is the final Steiner tree where $\cE_\cS$ only includes ``active" cross-cell edges in $\cG$, 
and edges identified in Step 5. 
\State Compute $D(\cG_\cS)=\sum_{e\in\cE_\cS}d_\cS(e)$.
\end{algorithmic}
\end{algorithm}

Voronoi cell computation 
naturally inclines to vertex-centric parallel processing~\cite{Gonzalez:2014:GGP:2685048.2685096}. 
Computation of a single cell closely resembles single source shortest path (SSSP) computation and Bellman-Ford based fast, vertex parallel SSSP algorithms are well known~\cite{Gonzalez:2014:GGP:2685048.2685096, Aasawat.BigData.2020.9378430}. It is worth noting that Ceccarello et al.~\cite{Ceccarello.2016.7515997} used the work-efficient $\Delta$-Stepping algorithm in parallel shortest path computation from multiple sources for diameter approximation of weighted graphs (comparable to Voronoi cell computation). We, however, base our distributed implementation of Voronoi cell computation on Bellman-Ford's algorithm which can harness asynchronous distributed processing (i.e., overlapping latency prone distributed communication with computation) and accelerate the convergence rate (details in \S\ref{section:system_description}). $\Delta$-Stepping is an iterative algorithm. Recently, Wang et al.~\cite{Wang.2021.ppopp.10.1145/3437801.3441605} presented a shared memory, asynchronous SSSP implementation for a single Nvidia GPU; the solution is an adaptation of the $\Delta$-Stepping algorithm. The technique, however, does not naturally extend to distributed memory. Our parallel algorithm computes all Voronoi cells in parallel -- comparable to running $|\cS|$ parallel instances of a parallel SSSP algorithm on the same graph. Alg.~\ref{alg:parallel_steiner_tree} presents a high-level overview of the proposed parallel algorithm and Fig.~\ref{fig:asmt_algorithm_steps_0011} illustrates the key algorithmic steps.   

Alg.~\ref{alg:parallel_steiner_tree} {\bf Step 1.} computes Voronoi cells in parallel based on Bellman-Ford's SSSP algorithm. For given $\cG$ and $\cS$, Fig.~\ref{alg:parallel_steiner_tree}(a) shows the computed Voronoi cells. Cross-cell edges, edges connecting two separate cells, are shown using broken lines. 
{\bf Step 2.} constructs the distance graph $\cG'_1$. Two cells can have multiple cross-cell edges, as shown in Fig.~\ref{alg:parallel_steiner_tree}(b). A parallel routine identifies unique min-distance edges, each connecting a pair of cells (a tie-breaking scheme is used to ensure uniqueness). The distance information of these cross-cell edges determines the edge weights of $\cG'_1$. 
{\bf Step 3.} generates $\cG'_2$, an MST of $\cG'_1$ (Fig.~\ref{alg:parallel_steiner_tree}(c)). Typically $\cG'_1$ is small, has at most ${|\cS| \choose 2}$ edges. Even for $|\cS|=10$K, there are less than 50M edges, orders of magnitude smaller than the billion-edge data graphs we used for evaluation in \S\ref{section:evaluation}. We argue that a sequential routine is sufficient for computing the MST $\cG'_2$. Given the problem size is often small, attempting to parallelize MST computation likely to yield marginal or no gain at all. This design choice is congruent with Bader et al.'s parallel MST approach~\cite{Bader.IPDPS.2004.1302953}: when the problem size becomes small, they switch to a sequential algorithm for the remainder of MST computation. In \S\ref{section:evaluation}, we show, for $|\cS|=10$K, the MST $\cG'_2$ can be computed in about two seconds, using a sequential C++ implementation of Prim's MST algorithm, often insignificant compared to total runtime. 
{\bf Step 4.} in parallel, marks all cross-cell edges in $\cG$ as ``deleted" except the ones whose corresponding edges are present in the MST $\cG'_2$. Fig.~\ref{alg:parallel_steiner_tree}(d) shows the deleted edges using light grey outline. 
{\bf Step 5.} in parallel, starting from each vertex $u, v$ of each ``active" cross-cell edge (shown using blue outline in Fig.~\ref{alg:parallel_steiner_tree}(d)), identifies all the edges in the shortest path from $u \in N(s) $ to $s \in \cS$ by following predecessor vertices identified during Voronoi cell computation. The remaining cross-cell edges after edge pruning in Step 4. and the edges identified in Step 5. form a valid Steiner tree, as shown in Fig.~\ref{alg:parallel_steiner_tree}(e). 

The 2-approximation bound $D(\cG_\cS)/D_{min}(\cG)\leq2(1-1/l)$ of our algorithm is guaranteed by Mehlhorn's proof~\cite{MEHLHORN1988125} that every MST of $\cG'_1$ (defined in \S\ref{section:preliminaries} and Alg.~\ref{alg:parallel_steiner_tree}) is also a MST of $\cG_1$ in Alg.~\ref{alg:kmb} which allows the 2-approximation bound of the KMB algorithm to apply to the algorithm in Alg.~\ref{alg:parallel_steiner_tree}.

\textbf{\textit{Time Complexity.}}
We briefly discuss sequential time complexity of key steps in Alg.~\ref{alg:parallel_steiner_tree}:  
Voronoi cell computation based on Bellman-Ford's algorithm requires $O(|\cV||\cE|)$ time. Identifying min-distance, cross-cell edges during the construction of $\cG'_1$ requires $O(|\cE|)$ time. The time complexity of finding the MST $\cG'_2$ using Prim's algorithm is $O(|\cE'_1|log|\cS|)$. Post MST edge pruning and identification of final Steiner tree edges, each requires no more than $O(|\cE|)$ time. 

\section{Distributed Implementation}
\label{section:system_description}

We present an MPI-based distributed implementation of the parallel Steiner tree algorithm introduced in \S\ref{section:solution_approach}. To accommodate large-scale graphs with 10s and 100s billion edges, we embrace a scale-out design: the data graph is partitioned; partitions have approximately equal share of vertices; each partition is assigned to an MPI process which enables processing the partitions in parallel. The solution employs a combination of vertex and edge centric processing, asynchronous processing 
and a message prioritization scheme towards offering fast time-to-solution and resources efficiency.

As highlighted in the time complexity analysis (\S\ref{section:solution_approach}), Voronoi cell computation is the most expensive step in our algorithm, therefore, we seek a solution that accelerates the throughput of this step. Previous studies~\cite{Xie.PPoPP.2015.10.1145/2858788.2688508, Aasawat.BigData.2020.9378430} showed that asynchronous processing offers notable advantage over bulk synchronous processing (BSP) for distributed shortest path computation: the former enabling faster convergence. To this end, we begin with HavoqGT~\cite{Pearce:2014:FPT:2683593.2683654} as the foundation and implement other features required by Alg.~\ref{alg:parallel_steiner_tree}. Our choice for HavoqGT is motivated by a number of considerations: HavoqGT, \textit{(i)} supports asynchronous processing, where latency prone communication can be overlapped with computation; \textit{(ii)} offers load balancing for scale-free graphs through vertex-cut partitioning by distributing edges of high-degree vertices across multiple partitions -- crucial to scale to large graphs with skewed degree distribution; and \textit{(iii)} an MPI-based implementation is likely more efficient than a Hadoop/Spark based solution~\cite{Gittens:2018:ALD:3219819.3219927}.
Note that our solution can easily be implemented by extending other general purpose graph processing frameworks that expose a vertex-centric API, e.g., Gluon-Galois~\cite{Dathathri.2019.PCT.8891625},  
GraphLab,  
Giraph, 
and GraphX~\cite{Gonzalez:2014:GGP:2685048.2685096}, independent of BSP or asynchronous processing approaches.

In HavoqGT, algorithms are implemented as vertex-callbacks: the user-defined $visit()$ callback accesses and updates a vertex's state(s). HavoqGT generates events, called a VISITOR, that invoke $visit()$ callbacks -- either on all graph vertices using the \mbox{$do\_traversal(arguments, traversal\_type \leftarrow init\_all)$} method, or for a neighboring vertex using the $push($VISITOR$)$ method. This enables asynchronous 
data exchange between graph vertices. 
The graph computation completes when all visitors in the message queue have been processed~\cite{Pearce:2014:FPT:2683593.2683654}.

In a distributed setting, each process runs an instance of Alg.~\ref{alg:distributed_steiner_tree} 
which iterates over a number of steps to produce the final 2-approximation Steiner minimal tree. The \mbox{INITIALIZATION} procedure allocates per-partition memory for distributed vertex and edge states and initializes them. For example, the distance graph $\cG'_1$, with $|\cS| \choose 2$ edges (with weights initialized to $\infty$), is created in this step (line 2); the map $\cE_N$, initially empty, identifies the cross-cell edges in $\cG$, is defined here (line 4). The computation steps under the STEINER\_TREE procedure can be categorized in to vertex and edge centric procedures; we describe them next.

\begin{algorithm}[!t]
\small
\caption{Distributed Steiner Tree Algorithm}
\label{alg:distributed_steiner_tree}
\hspace*{\algorithmicindent}\textbf{Input:} edge-weighted graph $\cG$, seed vertices $\cS$ \\
\hspace*{\algorithmicindent}\textbf{Output:} Steiner tree $\cG_\cS$, total distance $D(\cG_\cS)$
\begin{algorithmic}[1]
\Procedure{initialization}{}
\State $\cG'_1(\cV'_1=\cS,\cE'_1={\cS \choose 2}, d'_1=\infty)$, where $(s,t) \in \cE'_1 : s < t$ 
\State $\cG_\cS(\cV_\cS=\emptyset,\cE_\cS=\emptyset, d_\cS)$
\State $\cE_N \leftarrow \emptyset$  
\Comment map of cross-cell edges that maps $(s, t) \in \cE'$ to $(u, v) \in \cE$, where $u \in N(s)$, $v \in N(t)$, $u < v$    
\State $d_N$ is the distance function for keys (i.e., $(s, t) \in \cE'$) of $\cE_N$ 
\ForAll {$v\in\cV$}
  \If {$v\in\cS$}
    \State $src(v) \leftarrow v$; $pred(v) \leftarrow v$; $d_1(src(v), v)\leftarrow0$
  \Else
    \State $src(v) \leftarrow \infty$; $pred(v) \leftarrow \infty$; $d_1(src(v), v) \leftarrow \infty$
\EndIf
\EndFor
\EndProcedure

\Procedure{steiner\_tree}{}
\State VORONOI\_CELL\_ASYNC($\cG, \cS$); \textbf{barrier}
\State LOCAL\_MIN\_DIST\_EDGE\_ASYNC($\cG, \cS, \cE_N$); \textbf{barrier}
\State GLOBAL\_MIN\_DIST\_EDGE\_COLL($\cG, \cS, \cE_N$); \textbf{barrier}
\ForAll {$(s,t)\in\cE'_1$} \Comment partition local operation
\State $d'_1(s,t)\leftarrow d_N(s,t)$ 
\Comment{$d_N$ is set in Alg.~\ref{alg:min_weight_edge}, line 3}
\EndFor
\State $\cG'_2 \leftarrow$ MST\_SEQUENTIAL($\cG'_1$); \textbf{barrier}
\State EDGE\_PRUNING\_COLL($\cE_N$, $\cG'_2$); \textbf{barrier}
\State TREE\_EDGE\_ASYNC($\cG, \cE_N, \cG_\cS$); \textbf{barrier}
\State \Return $\cG_\cS$, $D(\cG_\cS)=\sum_{e\in\cE_\cS}d_\cS(e)$
\EndProcedure

\end{algorithmic}
\end{algorithm}

\textbf{Voronoi Cells} are computed asynchronously, initiated by invoking $do\_traversal()$. Alg.~\ref{alg:voronoi_cell} presents the corresponding $visit()$ procedure in HavoqGT's vertex-centric abstraction; lines 15--18 list the VISITOR states. Initially only a vertex $s \in \cS$ visits its neighbors (line 5). In general, when a vertex $v_j : src(v_j) = s$ is visited by another vertex $v_p : src(v_p) = t$, $s \neq t$, if $d_1(s, v_j) > d(v_j, v_p) + d_1(t, v_p)$ then $v_j$, becomes a member of the Voronoi cell $N(t)$, adopts $v_p$ as its predecessor, updates its distance accordingly (line 8), and notifies its neighbors about the state update (line 11). 

To accelerate convergence of shortest path computations in Voronoi cells, we employ a message prioritization technique in the message queue of each graph partition ($vq$ in line 3) which gives precedence to a message from a vertex at a lower distance. This can produce similar effect of the min-priority queue in Dijkstra's algorithm, enabling the Bellman-Ford-based distributed shortest path kernel to potentially converge faster. The technique is light-weight and best-effort only; its effectiveness depends on timeliness of asynchronous message propagation within the network which is nondeterministic. The worst-case complexity is the same as that of Bellman-Ford’s algorithm, however, even with message prioritization, it is not guaranteed the best-case complexity will match that of Dijkstra’s algorithm. In \S\ref{section:evaluation_priority_queue}, we demonstrate the benefits of using a priority message queue over a FIFO queue.

\begin{algorithm}[!t]
\small
\caption{Voronoi Cell}
\label{alg:voronoi_cell}
\begin{algorithmic}[1]

\Procedure{voronoi\_cell\_async ($\cG, \cS$)}{}
\State $do\_traversal(r \leftarrow 0, traversal\_type \leftarrow init\_all)$ 
\EndProcedure

\Procedure{visit}{$\cG, \cS, vq$}
\Comment $vq$ - (priority) message queue

\State $do\_update \leftarrow false$
\If{$d_1(src(v_j), v_j) = 0$ and $d_1(src(v_j), v_j) = r$}
\State $t \leftarrow src(v_j)$; $r \leftarrow d_1(src(v_j), v_j)$; $do\_update \leftarrow true$
\ElsIf {$0 < r < d_1(src(v_j),v_j)$}
\State $src(v_j) \leftarrow t$; $pred(v_j) \leftarrow v_p$; 
$d_1(src(v_j), v_j) \leftarrow r$
\State $do\_update \leftarrow true$
\EndIf

\If{$do\_update = true$}
\ForAll{$v_i \in adj(v_j)$}
\State $r \leftarrow r + d(v_i, v_j)$
\State $vq.push($VORONOI\_CELL\_VISITOR$(v_i, v_j, t, r))$
\EndFor
\EndIf
\EndProcedure

\State VORONOI\_CELL\_VISITOR
\State $v_j$ -- vertex that is being visited
\State $v_p$ -- vertex that sent the visitor to $v_j$
\State $t$ -- current source vertex of $v_p$
\State $r$ -- current distance of $v_p$ from $t$

\end{algorithmic}
\end{algorithm}

\textbf{Min Distance Edges} between Voronoi cells are identified towards creating the distance graph $\cG'_1$. First, the \mbox{LOCAL\_MIN\_DIST\_EDGE\_ASYNC} (Alg.~\ref{alg:min_weight_edge}) procedure identifies min-distance, cross-cell edges local to a partition and add them to the local copy of $\cE_N$. This is an asynchronous procedure based on HavoqGT's vertex-centric abstraction. The distance information of a vertex computed in Alg.~\ref{alg:voronoi_cell} is stored locally. To compute $d_N(s,t)$ (Alg.~\ref{alg:min_weight_edge}, line 3), a vertex $u$ needs to receive $d_1(v, t)$ from $v$, possibly from a remote partition; hence a distributed routine is required. GLOBAL\_MIN\_DIST\_EDGE\_COLL (Alg.~\ref{alg:min_weight_edge}) is an edge-centric procedure: it performs MPI\_ALLreduce(MPI\_MIN) collective operation on distance values of local copies of $\cE_N$ to identify global min-distances. 
$\cG'_1$ is updated with the distance information in the global $\cE_N$ (Alg.~\ref{alg:distributed_steiner_tree}, line 15).

\begin{algorithm}[!t]
\small
\caption{MIN Distance Edge and Global Edge Pruning}
\label{alg:min_weight_edge}
\begin{algorithmic}[1]

\Procedure{local\_min\_dist\_edge\_async ($\cG, \cS, \cE_N$)}{}
\State for every $(u, v) \in \cE$ local to a partition of $\cG$, computes
\State $d_N(s, t) = min(d_1(s, u) + d(u, v) + d_1(v, t))$, where
\State $u \in N(s)$, $v \in N(t)$, $s \neq t$; adds each corresponding 
\State min-distance edge $(u, v)$ the local map, 
\State i.e., $\cE_N:(s, t) \leftarrow (u, v)$ 
\EndProcedure

\Procedure{global\_min\_dist\_edge\_coll ($\cG, \cS, \cE_N$)}{}
\State performs MPI\_Allreduce(MPI\_MIN) on local distance  
\State values (i.e., $d_N(s, t)$) of $\cE_N$ and save results in global $\cE_N$
\EndProcedure

\Procedure{edge\_pruning\_coll ($\cE_N, \cG'_2$)}{}
\State removes all cross-cell edges from the global $\cE_N$     
\State whose corresponding edges are not present in $\cG'_2$;   
\State performs MPI\_Allreduce(MPI\_MIN) on source vertex \State IDs of edges, i.e., $(u, v) \rightarrow (s,t) \in \cE_N$ (global copy),   
\State to ensure only one cross-cell edge per Voronoi cell pair
\EndProcedure

\end{algorithmic}
\end{algorithm}

\textbf{MST} $\cG'_2$ of the distance graph $\cG'_1$ is computed using a sequential routine (Alg.~\ref{alg:distributed_steiner_tree}, line 17); our current implementation uses Boost's implementation of Prim's algorithm. Since $\cG'_1$ has only $|\cS| \choose 2$ edges, it is replicated on all partitions. Partitions locally compute $\cG'_2$ and avoid remote memory copy operations.         

\textbf{Global Edge Pruning} is a distributed edge-centric routine  (Alg~\ref{alg:min_weight_edge}): 
First, it removes the edges from global $\cE_N$, whose corresponding edges are not present in $\cG'_2$. Then, it performs collective operation MPI\_Allreduce(MPI\_MIN) on source vertex IDs of edges in $\cE_N$ to ensure only a unique cross-cell edge exists for each unique pair of Voronoi cells (multiple cross-cell edges with identical distance can bridge the same two cells).

\textbf{Steiner Tree Edges} local to each Voronoi cell are identified by a vertex-centric asynchronous routine (Alg.~\ref{alg:tree_edge}): starting from each vertex $u, v$ of every cross-cell edge $(u, v) \in \cE$ present in $\cE_N$, tree edges within their respective Voronoi cells are identified by recursively visiting predecessors until the source vertex, e.g., $s : u \in N(s)$, has been reached. Note that the cross-cell edges in $\cE_N$ also belong to the final Steiner tree $\cG_\cS$ (line 4). Alg.~\ref{alg:tree_edge} significantly reduces the number of messages communicated since often the number of Steiner tree edges $|\cE_\cS|$ is orders of magnitude smaller than number of non-tree edges $|\cE|-|\cE_\cS|$ (see Table~\ref{table:evaluation_tree_vertices} for empirical evidence).

\begin{algorithm}[!h]
\small
\caption{Steiner Tree Edge}
\label{alg:tree_edge}
\begin{algorithmic}[1]
\Procedure{tree\_edge\_async ($\cG, \cE_N, \cG_\cS$)}{}
\ForAll{$(u, v) \rightarrow (s,t) \in \cE_N$}
\Comment defined in Alg.~\ref{alg:distributed_steiner_tree}, line 4
\If{this partition of $\cG$ is $u$'s home partition}
\State $\cE_\cS \leftarrow \cE_\cS \cup  (u, v)$; 
$d_\cS(u, v) \leftarrow d(u, v)$
\State $vq.push($TREE\_EDGE\_VISITOR$(u))$ 
\State $vq.push($TREE\_EDGE\_VISITOR$(v))$         
\EndIf
\EndFor
\State $do\_traversal()$ 
\EndProcedure

\Procedure{visit}{$\cG, \cG_\cS, vq$}

\If{$v_j \neq src(v_j)$ }
\State $\cE_\cS \leftarrow \cE_\cS \cup  (pred(v_j), v_j)$
\State $d_\cS(pred(v_j), v_j) \leftarrow d(pred(v_j), v_j)$
\If{$pred(v_j) \neq src(v_j)$}
\State $vq.push($TREE\_EDGE\_VISITOR$(pred(v_j))$
\EndIf
\EndIf
\EndProcedure

\State TREE\_EDGE\_VISITOR
\State $v_j$ -- vertex that is being visited

\end{algorithmic}
\end{algorithm}

\section{Evaluation}
\label{section:evaluation}

We present strong scaling experiments using billion-edge, real-world graphs and up to 512 compute nodes (\S\ref{section:evaluation_strong_scaling}); demonstrate support for thousands of seed vertices (\S\ref{section:evaluation_terminal_vertex}); evaluate the effectiveness of our design choices and optimizations (\S\ref{section:evaluation_priority_queue}); study influence of problem artifacts on performance and sensitivity of our solution to problem parameters (\S\ref{section:evaluation_edge_weight}, \S\ref{section:evaluation_seed_selection} and \S\ref{section:evaluation_memory_usage}); compare performance of our solution with related work and measure result quality (\S\ref{section:evaluation_compare_related_work}).        

\textbf{\textit{Testbed.}}
The testbed is a 2.6 petaFLOP cluster comprised of over 2K compute nodes and the Intel Omni-Path interconnect. Each node has two 18-core Intel Xeon E5-2695v4 @2.10GHz processors and 128GB of memory~\cite{quartz.001}. 
We run 16 MPI processes per node (as observed, this configuration offers the best performance, since each process runs two threads). 

\textbf{\textit{Graph Datasets.}} Table~\ref{table:evaluation_datasets} summarizes the main characteristics of the real-world datasets used for evaluation and shows their storage requirements in HavoqGT binary graph format. 
For each graph, we create symmetric edges ($2|\cE|$ edges) with non-zero, positive edge weights in the range as listed in Table~\ref{table:evaluation_datasets}. The Web Data Commons 2012 (WDC), ClueWeb 2012 (CLW), and UK Web 2007-05 (UKW) are web graphs whose vertices are webpages and edges are hyperlinks~\cite{Dathathri.2019.PCT.8891625}. Friendster (FSR)~\cite{Dathathri.2019.PCT.8891625} and LiveJournal (LVJ)~\cite{Reza.2020.SIGMOD.10.1145/3318464.3380566} are user-centric social media  graphs. 
Patent (PTN) and CiteSeer (CTS) are citation graphs~\cite{Reza.2020.SIGMOD.10.1145/3318464.3380566}. MiCo (MCO) is a co-author graph created from the Microsoft research article repository~\cite{Reza.2020.SIGMOD.10.1145/3318464.3380566}. 

\begin{table}[!t]
\footnotesize
\renewcommand{\arraystretch}{1.2}
\captionsetup{font=footnotesize}
\caption{Characteristics of graph datasets used for evaluation.}
\label{table:evaluation_datasets}
\setlength{\tabcolsep}{3.0pt}
\centering
\begin{tabular}{lrrrrrrr}
\hline
& 
\multicolumn{1}{c}{$|\cV|$} & \multicolumn{1}{c}{$2|\cE|$} & \multicolumn{1}{c}{Max.} & \multicolumn{1}{c}{Avg.} & \multicolumn{1}{c}{Edge} & \multicolumn{1}{c}{Size} \\
& & & \multicolumn{1}{c}{degree} & \multicolumn{1}{c}{degree} & \multicolumn{1}{c}{weight} & \\
\hline
WDC12
& 3.5B & 257B & 95M & 72.3 
& [1, 500K]
& 5.7TB \\ 
ClueWeb12
& 978M & 85B & 75.6M & 87 
& [1, 100K]
& 1.9TB \\ 
UKWeb07
& 105M & 7.5B & 975K & 71 
& [1, 75K]
& 150GB \\ 
Friendster
& 66M & 3.6B & 5.2K & 55.1 
& [1, 50K]
& 84GB \\ 
LiveJournal
& 4.8M & 85.7M & 20.3K & 17.7 
& [1, 5K]
& 2.1GB \\
Patent
& 2.7M & 28M & 789 & 10.2 
& [1, 5K]
& 692MB \\
MiCo
& 100K &  2.2M & 1.4K & 22 
& [1, 2K]
& 52MB \\
CiteSeer
& 3.3K &  9.4K & 99 & 3.6 
& [1, 1K]
& 328KB \\
\hline
\end{tabular}
\end{table}

\textbf{\textit{Seed Vertex Selection.}} To ensure all seed vertices are present in the Steiner tree, first, we identify the largest connected component using Breath-first search (BFS) and BFS-levels of vertices. It is undesirable that majority of seed vertices are directly connected in which case Voronoi cell computation could converge faster. To avoid this scenario, from different BFS-levels, we randomly select vertices -- often a higher percentage of vertices are selected from a level with higher vertex frequency. Note that the length of the weighted shortest path between a vertex pair is unlikely to be the same as in the BFS-tree. Also, sampling disjoint sets by the population size roughly converges to uniform random overall.

\begin{figure*}[!t]
\centering
\includegraphics[width=\linewidth]{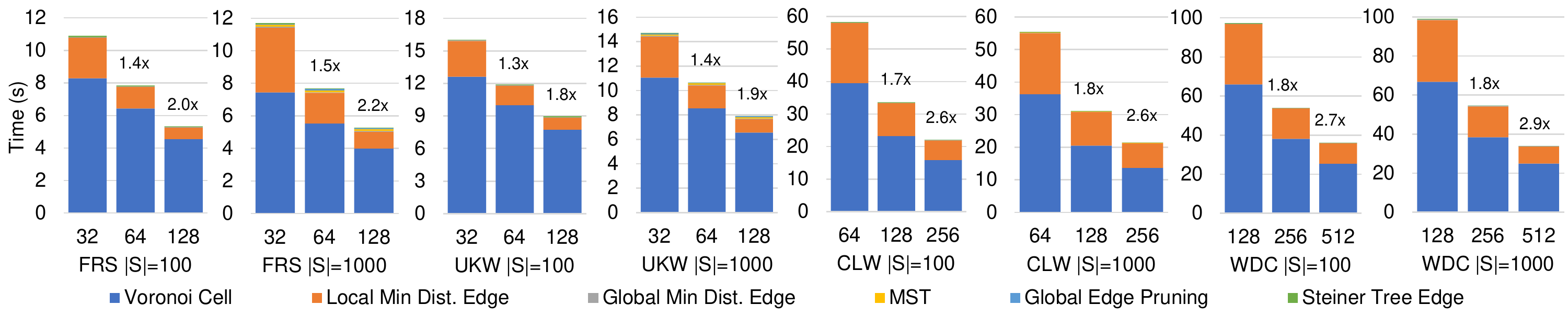} 
\setlength{\belowcaptionskip}{-12pt}
\captionsetup{font=footnotesize}
\caption{Strong scaling experiment results using the FRS, UKW, CLW and WDC datasets. Runtime is broken down to the key computation steps (identified in chart legends). For each experiment, X-axis labels in the first row indicates the platform size (number of compute nodes). Speedup achieved over the smallest scale is shown on top of respective stacked bar plots.}
\label{fig:asmt_strong_scaling_0011}
\end{figure*}

\textbf{\textit{Experiment Methodology.}} The performance metric is the time-to-solution, i.e., identifying the edges in a Steiner tree. It does not including graph partitioning and loading times. Runtime numbers are averages over at least 10 runs. For strong scaling experiments, the smallest platform scale is the one that can accommodate a dataset and required algorithm states. Strong scaling experiments do not include results on a single node since this does not involve remote communication. 

\subsection{Strong Scaling}
\label{section:evaluation_strong_scaling}
Fig.~\ref{fig:asmt_strong_scaling_0011} presents results of strong scaling experiments using, the four largest graphs in Table~\ref{table:evaluation_datasets}, 100 and 1K seed vertices, and up to 512 compute nodes. Runtime is broken down to the key computation steps described in \S\ref{section:system_description}. For each dataset, the smallest platform scale (e.g., 32 nodes for FRS) is the one that can accommodate the graph and required algorithm states (each machine has 128GB memory). For all datasets, majority of the runtime is spent in computing Voronoi cells, followed by local min-distance edge computation whose scalability improves almost linearly when the platform size is doubled. More time is spent in local min-distance edge computation when the number of seeds is increased by an order of magnitude. 
Time spent in the remaining distributed computation phases are insignificant compared to the total runtime (hence, scalability). Moving from 100 to 1K seeds, for a dataset, no major deviation is observed in scaling performance. Overall, we observe decent scaling -- the larger CLW and WDC graphs demonstrate better scaling (up to 90\% efficient scaling) compared to the smaller FRS and UKW datasets. Voronoi cell computation, whose performance is most affected by the irregular graph topology, is the main scalabaility bottleneck.      

\begin{figure}[!t]
\centering
\includegraphics[width=2.8in]{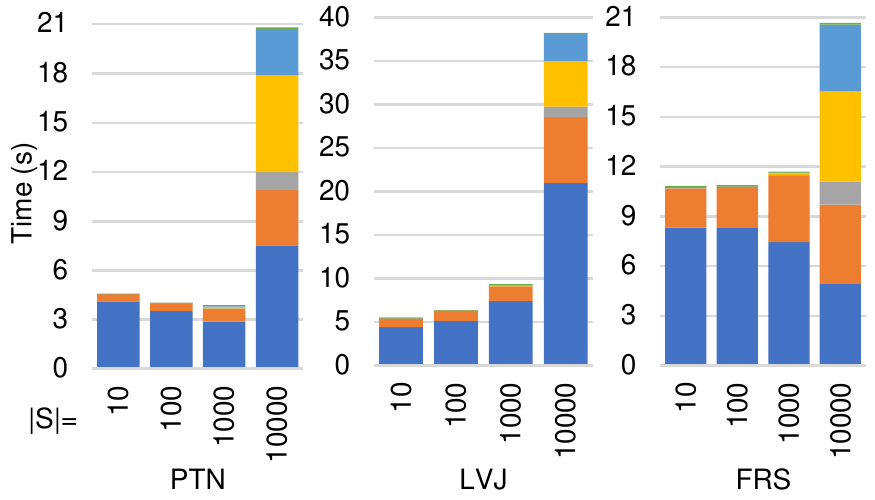} 
\includegraphics[width=2.8in]{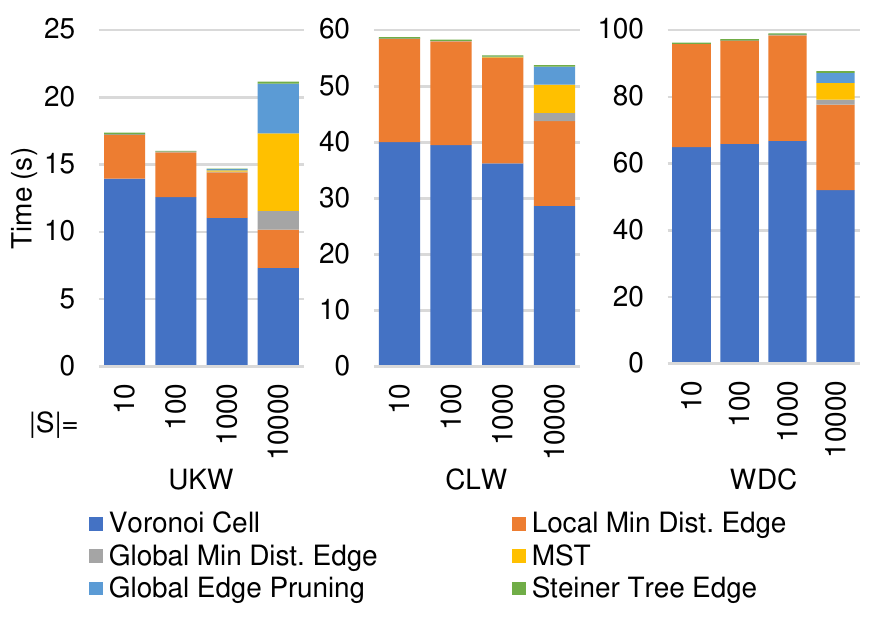} 
\setlength{\belowcaptionskip}{-12pt}
\captionsetup{font=footnotesize}
\caption{Runtime performance of different seed vertex counts $|\cS|$: 10, 100, 1K and 10K, using six graph datasets. For each dataset, the number of processes used are the same for all values of $|\cS|$. For $|\cS|=$ 10K, we had to double the number of compute nodes to allow more memory for the MPI collective operation on buffers with $\sim$50M items (i.e., edges of the distance graph $\cG'_1$).}
\label{fig:asmt_source_scaling_0011}
\end{figure}

\subsection{Number of Seed Vertices vs Runtime Performance}
\label{section:evaluation_terminal_vertex}
\begin{figure*}[!t]
\centering
\subfloat[$|\cS|$=10]{\includegraphics[width=2.2in]{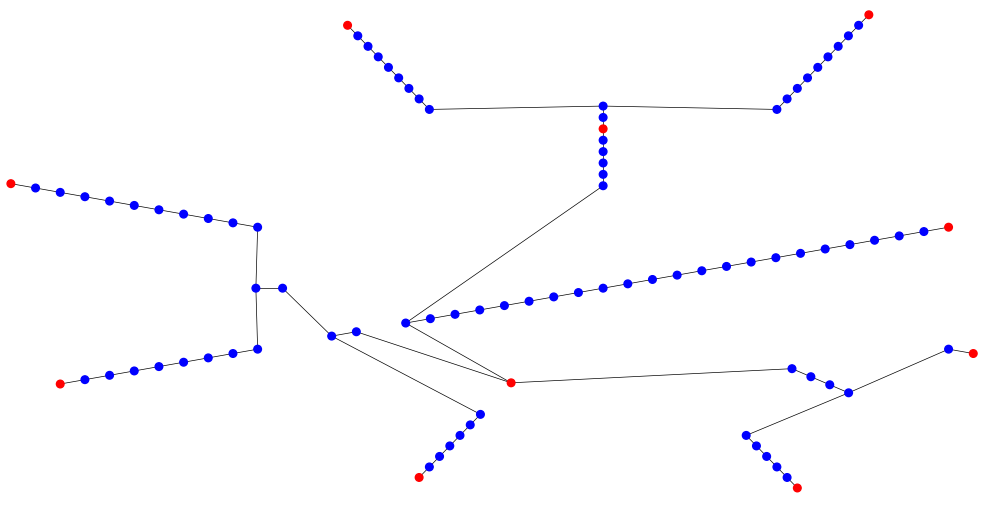}}
\subfloat[$|\cS|$=100]{\includegraphics[width=2.3in]{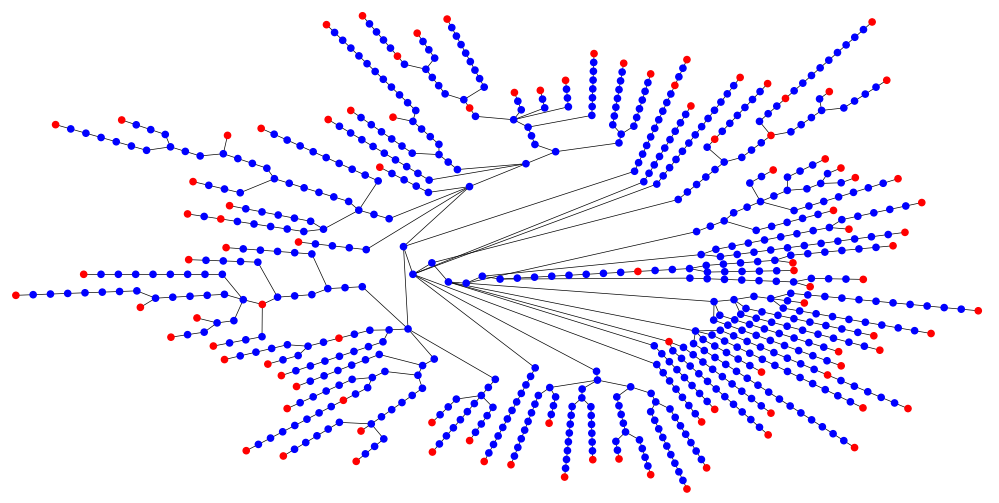}} 
\subfloat[$|\cS|$=1000]{\includegraphics[width=2.6in]{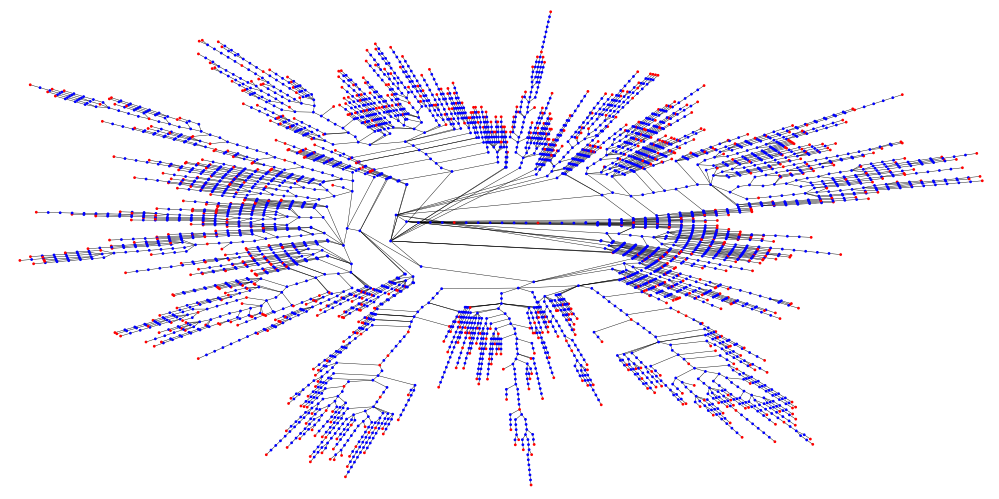}}
\setlength{\belowcaptionskip}{-12pt}
\captionsetup{font=footnotesize}
\caption{Steiner trees in the MiCo graph for three seed sets of different sizes. Seed vertices have red fill and Steiner vertices have blue fill.}
\label{fig:asmt_mico_output_0011}
\end{figure*}

Fig.~\ref{fig:asmt_source_scaling_0011} compares runtime performance of different seed vertex counts $|\cS|$: 10, 100, 1K and 10K, using six graph datasets. For each dataset, the number of processes used are the same for all values of $|\cS|$. Runtime is broken down to the key computation steps described in \S\ref{section:system_description}. Except for smaller PTN and LVJ, Voronoi cell computation time decreases for the highest value of $|\cS|$ (i.e., 10K). This is due to the convergence rate being accelerated in the presence of a large $\cS$ (a data dependent artifact). For $|\cS|=$ 10, 100 and 1K, time spent in the final four steps (i.e, global min-distance edge, MST, global edge pruning and Steiner tree edge) is insignificant compared to the total runtime. For $|\cS|=$ 10K, the distance graph $\cG'_1$ has $\sim$50M edges -- two orders of magnitude larger than the distance graph for $|\cS|=$ 1K. Updating the edge-weights, computing MST on the $\sim$50M edge graph and using this information for further edge pruning accounts for some computation time. The influence is more visible in smaller graphs. Note that reported time for the MST step also includes time spent in moving results from the sequential code to the distributed data structure; for $|\cS|=$ 10K, the actual spanning tree computation takes $\sim$2s only. Table~\ref{table:evaluation_tree_vertices} lists the number of edges in the final tree for each dataset and seed vertex combination. Fig.~\ref{fig:asmt_mico_output_0011} shows Steiner trees in the MiCo graph for the given seed sets. 

\begin{table}[!h]
\footnotesize
\renewcommand{\arraystretch}{1.2}
\captionsetup{font=footnotesize}
\caption{Total number of edges in the output Steiner tree for different graphs and seed vertex sets. 
}
\label{table:evaluation_tree_vertices}
\setlength\tabcolsep{3.0pt}
\centering
\begin{tabular}{r r r r r r r r r}
\hline
$|\cS|$ & WDC & CLW & UKW & FRS & LVJ & PTN & MCO & CTS \\
\hline
10 & 326 & 152 & 184 & 149 & 105 & 125 & 93 & 66 \\
100 & 1,953 & 1,265 & 1,542 & 1,485 & 1,108 & 1,112 & 743 & 320 \\
1K & 12,488 & 6,909 & 10,644 & 11,639 & 7,193 & 8,075 & 4,599 & 1,362 \\
10K & 85,586 & 36,397 & 42,782 & 85,211 & 50,530 & 51,988 & N/A & N/A \\
\hline
\end{tabular}
\end{table}

\begin{figure}[!t]
\centering
\includegraphics[width=2.7in]{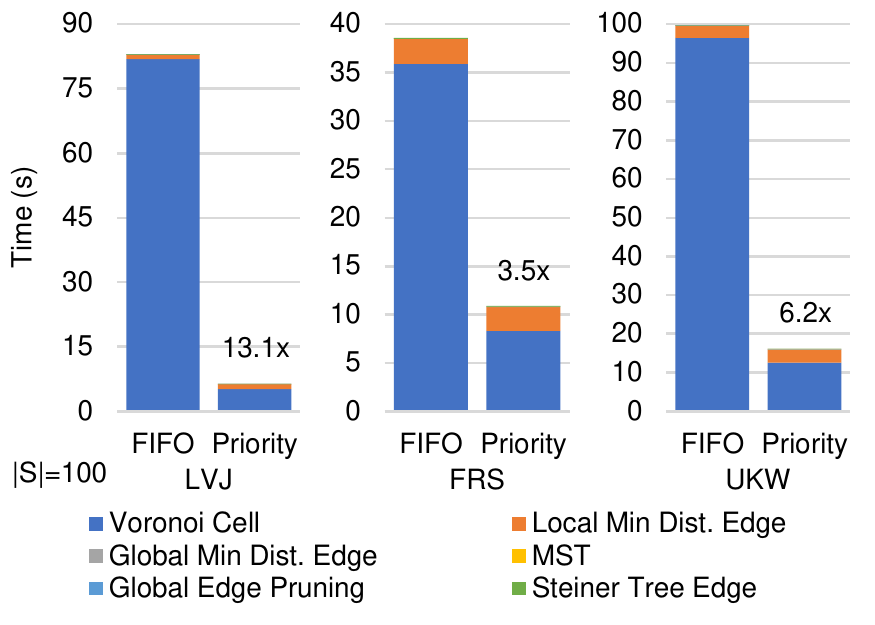} 
\setlength{\belowcaptionskip}{-12pt}
\captionsetup{font=footnotesize}
\caption{Runtime comparison of using FIFO and priority queues using a fixed number of seed vertices and cluster sizes (one node for LVJ, and 32 nodes for FRS and UKW). Runtime is broken down to the key computation steps (identified in chart legends). Speedup achieved due to the use of priority queue is shown on top of respective stacked bar plots.}
\label{fig:asmt_message_queue_0011}
\end{figure}

\begin{figure}[!t]
\centering
\includegraphics[width=2.7in]{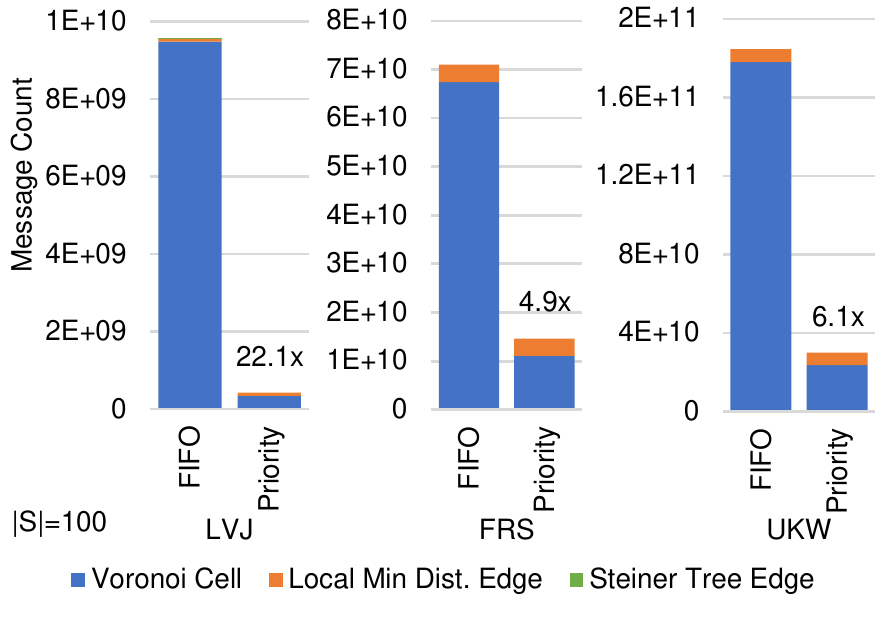} 
\setlength{\belowcaptionskip}{-12pt}
\captionsetup{font=footnotesize}
\caption{Comparison of message count due to the use of FIFO and priority queues for the same experiments in Fig.~\ref{fig:asmt_message_queue_0011}. Message count is broken down to the key computation steps (identified in chart legends). Improvement in generated message count due to the use of priority queue is shown on top of respective stacked bar plots. Note that the figure does not show message count for computation phases that rely on MPI collective operations (Alg.~\ref{alg:distributed_steiner_tree}).}
\label{fig:asmt_message_0011}
\end{figure}

\subsection{FIFO vs Priority Queue for Message Propagation}
\label{section:evaluation_priority_queue}
In \S\ref{section:system_description}, we discussed employing priority queue for message propagation within HavoqGT, specifically to accelerate the convergence of Voronoi cell computation. Here, we demonstrate the advantage of this design optimization with respect to runtime and message volume. Fig.~\ref{fig:asmt_message_queue_0011} compares runtime performance of using a priority queue with that of using a FIFO queue (default in HavoqGT). The advantage of priority queue is significant: for FRS, priority queue offers 3.5$\times$ speedup, while for LVJ, the improvement is over 13$\times$. Fig.~\ref{fig:asmt_message_0011} shows the actual number of messages communicated, grouped by computation phases, for the same set of experiments as in Fig.~\ref{fig:asmt_message_queue_0011}. These two set of results complement each other: improvement in runtime of Voronoi cell computation is a direct result of reduction in number of messages when the priority queue is used -- the improvement is 4.9$\times$ for FRS, while LVJ shows 22.1$\times$ improvement in generated message traffic. The local min-distance edge computation phase is responsible for a small portion of the total generated message traffic which is no greater than $|\cE|$. The number of messages due to the final Steiner tree edge identification phase is comparatively insignificant since the resulting Steiner tree is typically orders of magnitude smaller than the original data graph.

\subsection{Edge Weight Distribution vs Runtime Performance}
\label{section:evaluation_edge_weight}
\begin{figure}[!t]
\centering
\includegraphics[width=3.0in]{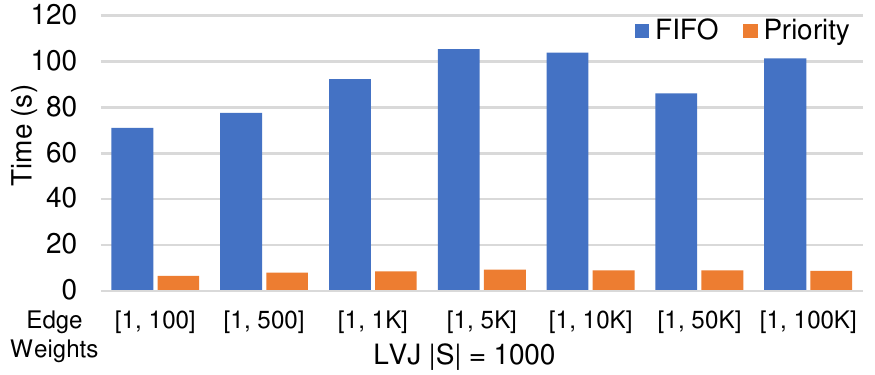} 
\setlength{\belowcaptionskip}{-12pt}
\captionsetup{font=footnotesize}
\caption{Influence of edge weight distribution on end-to-end runtime performance: X-axis labels are the edge weight range (inclusive). We compare two cases, using FIFO and priority queues for message propagation. Here, $\cS$ is fixed, 1K seeds, and the LVJ graph uses a single machine for all experiments.}
\label{fig:asmt_edge_weights_0011}
\end{figure}

Edge weight distribution is known to influence the convergence time of shortest path computation~\cite{Xie.PPoPP.2015.10.1145/2858788.2688508, Aasawat.BigData.2020.9378430}. Fig.~\ref{fig:asmt_edge_weights_0011} studies this phenomena using the LVJ graph for a fixed $\cS$ but different edge weight distributions, from [1, 100] up to [1, 100K]. We compare two cases, using FIFO and priority message queues. Edge distribution does impact runtime, especially Voronoi cell computation; for both message queues, edge weight distribution [1, 100] achieves the fastest convergence time. The variability in convergence time when using FIFO queue is much higher compared to using priority queue: for FIFO queue, the standard deviation is 13.5s, 14.7$\times$ higher than that of priority queue (which is only 0.91s). The results suggest the priority queue optimization makes our solution not only to perform better (for the LVJ graph, on average 10.8$\times$ faster than using FIFO queue) but also less sensitive to edge weight distribution. Note that these results are subjected to randomness associated with edge weight assignment.

\begin{table}[!b]
\footnotesize
\renewcommand{\arraystretch}{1.2}
\captionsetup{font=footnotesize}
\caption{Comparison of seed selection strategies using the LVJ graph. For each technique, the table lists runtime, number of the edges $|\cE_\cS|$ in the output Steiner tree $|\cG_\cS|$, and its total distance $D(\cG_\cS)$.}
\label{table:evaluation_seed_selection}
\setlength\tabcolsep{3.0pt}
\centering
\begin{tabular}{r | r r r | r r r}
\hline
& \multicolumn{3}{c|}{BFS-level} & \multicolumn{3}{c}{Uniform Random} \\ 
\hline
\multicolumn{1}{c|}{$|\cS|$} & \multicolumn{1}{c}{Time} & \multicolumn{1}{c}{$D(\cG_\cS)$} & \multicolumn{1}{c|}{$|\cE_\cS|$} & \multicolumn{1}{c}{Time} & \multicolumn{1}{c}{$D(\cG_\cS)$} & \multicolumn{1}{c}{$|\cE_\cS|$} \\
\hline
100 & 6.4s & 426.9K & 1,108 & 5.5s & 207.5K & 889 \\
1K & 9.3s & 2,840.9K & 7,193 & 9.0s & 1,845.5K & 7,202 \\
10K & 38.2s & 28,903.3K & 50,530 & 37.5s & 18,123.5K & 49,755 \\
\hline
& \multicolumn{3}{c|}{Eccentric} & \multicolumn{3}{c}{Proximate} \\ 
\hline
\multicolumn{1}{c|}{$|\cS|$} & \multicolumn{1}{c}{Time} & \multicolumn{1}{c}{$D(\cG_\cS)$} & \multicolumn{1}{c|}{$|\cE_\cS|$} & \multicolumn{1}{c}{Time} & \multicolumn{1}{c}{$D(\cG_\cS)$} & \multicolumn{1}{c}{$|\cE_\cS|$} \\
\hline
100 & 6.1s & 412.3K & 1,115 & 7.6s & 16.0K & 272 \\
1K & 6.3s & 6,091.5K & 6,548 & 8.1s & 101.0K & 1,699 \\
10K & 31.4s & 49,644.5K & 49,691 & 39.1s & 1,105.9K & 16,624 \\
\hline
\end{tabular}
\end{table}

\subsection{Studying Seed Selection Alternatives}
\label{section:evaluation_seed_selection}
Earlier in the section, we discussed the seed selection approach used for evaluating this work (we call it BFS-level). In this section, we explore three additional seed selection alternatives and study their influence on performance and characteristics of the output tree. Uniform random -- we randomly select $|\cS|$ vertices from the largest connected component. Eccentric -- we ensure seed vertices are faraway from each other. We use a technique inspired by the k-BFS heuristic~\cite{Iwabuchi.2018.cluster.8514886} -- the algorithm iteratively identifies ($k=|\cS|$) BFS sources which are then used as seeds for Steiner tree computation. Starting with a random vertex in the largest connected component as the BFS source, each of the subsequent $k-1$ BFS sources is identified based on BFS-levels computed in the previous $k-n$ rounds ($1 \leq n \leq k-1$): BFS source for the $k-n+1$'th round  $u_{k-n+1}=max(\{\sum_{j = 0}^{k-n} l_j(v_i)\})$, where $u_{k-n+1}, v_i \in \cV$, $u_{k-n+1} \neq v_i$, $l_j(v_i)$ is the BFS-level of $v_i$ in the $j$'th round. Proximate -- seed vertices are selected such that they are close to each other, following the same approach in the eccentric case: BFS source for the $k-n+1$'th round $u_{k-n+1}=min(\{\sum_{j = 0}^{k-n} l_j(v_i)\})$.  Table~\ref{table:evaluation_seed_selection} presents the results. In summery, we do not observe notable difference in performance between techniques. Compared to other techniques, proximate produces significantly smaller trees (which we tried to avoid in the evaluation of our work).

\subsection{Memory Usage Analysis}
\label{section:evaluation_memory_usage}
\begin{figure}[!t]
\centering
\includegraphics[width=2.7in]{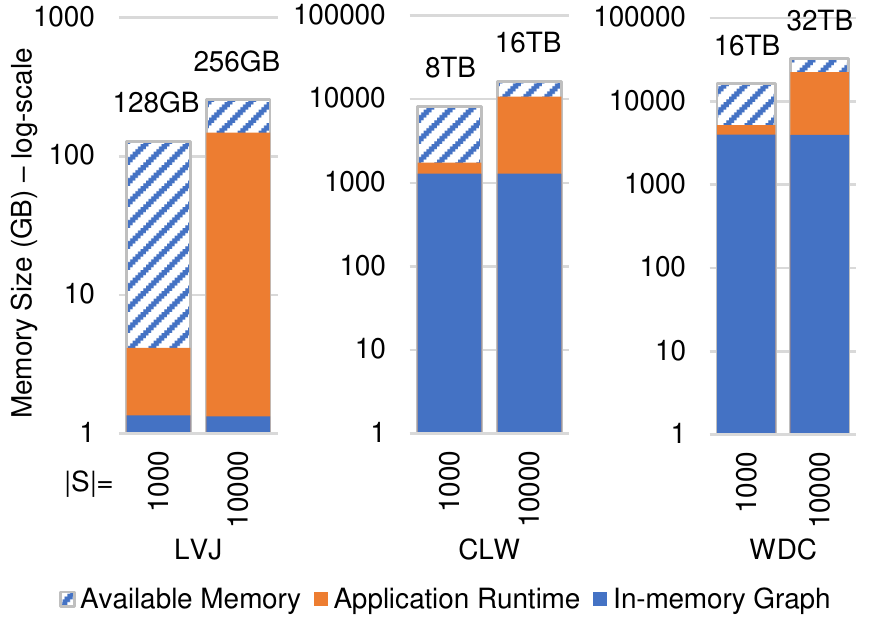} 
\setlength{\belowcaptionskip}{-12pt}
\captionsetup{font=footnotesize}
\caption{Cluster-wide peak memory usages by three graphs for 1K and 10K seed vertices. The total system memory is listed on top of respective stacked bars. For both $|\cS|=$ 1K and 10K experiments, the number of processes used are the same but the number of nodes is doubled for $|\cS|=$ 10K to allow more memory.}
\label{fig:asmt_memory_0011}
\end{figure}

Fig.~\ref{fig:asmt_memory_0011} shows cluster-wide peak memory usages by three graphs for 1K and 10K seed vertices. For both experiments, the process count is the same but the number of compute nodes is doubled in $|\cS|=$ 10K to meet the memory demand. $|\cS|=$ 1K experiments for LVJ, CLW and WDC use one, 64 and 128 nodes, respectively. For each experiment in Fig.~\ref{fig:asmt_memory_0011}, total memory usage is broken down to memory required for the in-memory HavoqGT binary graph and algorithm states (which includes communication buffers and messages). For the smaller LVJ, memory usage is dominated by algorithm states; $|\cS|=$ 10K consumes 35.9$\times$ more memory than $|\cS|=$ 1K. For the larger CLW and WDC, the differences are 4.4$\times$ and 1.7$\times$, respectively. Note that for $|\cS|=$ 10K, noticeable increase in memory usage is due to the MPI collective operation on the edge buffer $\cE_N$ which has $\sim$50M elements. Memory consumption improves when, instead of a single collective operatio on the entire edge buffer, multiple collective operations are performed on smaller chunks, e.g., 500K or 1M items per chunk, at the expense of runtime performance of course. 

\subsection{Comparison with Related Work}
\label{section:evaluation_compare_related_work}
We compare our distributed solution with two sequential 2-approximate algorithms WWW~\cite{Wu1986FA} and Mehlhorn~\cite{MEHLHORN1988125}, and sequential exact solution SCIP-JACK~\cite{Rehfeldt.2021.10.1007/978-3-030-55240-4_10}. ParaSCIP~\cite{GamrathKochRehfeldtetal.2014} enables SCIP-Jack to harness distributed clusters by replicating the data graph on all processes. Unfortunately, ParaSCIP did not yield any performance advantage in our experiments: the authors of SCIP-Jack reproduced our experiments and concluded that branch-and-bound search was lacking parallelism for the datasets we used (explained in~\cite{Rehfeldt.2021.10.1007/978-3-030-55240-4_10}); hence, we present results obtained using sequential SCIP-JACK 2.0 (reproduced by original authors as well). Unfortunately, we do not have access to another parallel, distributed implementation that solves the same problem 
that we could use for direct comparison. We implement WWW and Mehlhorn algorithms in C++ based on cache friendly CSR graph data structure. 
We use four small graphs in Table~\ref{table:evaluation_datasets} and up to 1K seed vertices. 

\begin{table}[!t]
\footnotesize
\renewcommand{\arraystretch}{1.2}
\captionsetup{font=footnotesize}
\caption{Runtime comparison between our distributed solution (D) -- using 16 processes on a single machine, and three related work: exact solution SCIP-Jack (S)~\cite{Rehfeldt.2021.10.1007/978-3-030-55240-4_10}, and 2-approximation algorithms WWW (W)~\cite{Wu1986FA} and Mehlhorn (M)~\cite{MEHLHORN1988125}. Time units: hour (h), minute (m), second (s), and millisecond (ms).}
\label{table:evaluation_comparison_related_work_runtime}
\setlength\tabcolsep{1.0pt}
\centering
\begin{tabular}{l|rrrr|rrrr|rrrr}
\hline
$|\cS|$ & \multicolumn{4}{c|}{10} & \multicolumn{4}{c|}{100} & \multicolumn{4}{c}{1000}\\ 
\hline
& S & W & M & D & S & W & M & D & S & W & M & D \\
\hline
LVJ & 9.4m & 27.8s & 25.1s & 5.5s & 10.1m & 28.4s & 40.5s & 6.4s & 45.8m & 28.4s & 1.9m & 9.3s \\
PTN & 7.3m & 8.4s & 14.8s & 4.6s & 11.0m & 8.3s & 28.7s & 4.0s & 1.0h & 8.4s & 1.5m & 3.9s \\
MCO & 8.8s & 0.3s & 0.2s & 0.3s & 9.8s & 0.3s & 0.4s & 0.3s & 53.5s & 0.3s & 1.6s & 0.5s \\
CTS & $<$1s & 0.5ms & 0.9ms & 5ms & $<$1s & 0.7ms & 4.9ms & 6ms & $<$1s & 2.2ms & 0.1s & 0.2s \\
\hline
\end{tabular}
\end{table}

Table~\ref{table:evaluation_comparison_related_work_runtime} presents comparison of the runtime performance. 
While for smaller CTS and MCO graphs, work efficient WWW performances slightly better, the advantage of our parallel solution becomes apparent for larger PTN and LVJ graphs: when using 16 processes on a single machine, our solution is maximum 27$\times$ faster than Mehlhorn and 5$\times$ faster than WWW. Table~\ref{table:evaluation_comparison_related_work_error} shows the quality of approximation of our distributed solution. 
We compare the total wight of a Steiner tree identified by our algorithm with that of the Steiner minimal tree produced by exact solution SCIP-Jack, i.e., we measure $D(\cG_\cS)/D_{min}(\cG)$. The table also lists the percentage error in approximation relative to the optimal solution. The empirical results complement the 2-approximation bound that our solution guarantees: on average, the total distance $D(\cG_\cS)$ of the Steiner tree identified by our solution is 1.0527 times greater than that of the Steiner minimal tree (i.e., $D_{min}(\cG)$) which is well within the theoretical bound of $\leq 2(1 - 1/l)$, and the approximation error is 5.3\%.

\begin{table}[!t]
\footnotesize
\renewcommand{\arraystretch}{1.2}
\captionsetup{font=footnotesize}
\caption{The quality of approximation of our distributed solution: on the left, the approximation ratio of a Steiner tree identified by our solution, i.e., $D(\cG_\cS)/D_{min}(\cG)$; on the right \% error of approximation for the same set of experiments. Steiner minimal trees were computed using SCIP-Jack~\cite{Rehfeldt.2021.10.1007/978-3-030-55240-4_10}.}
\label{table:evaluation_comparison_related_work_error}
\setlength\tabcolsep{3.0pt}
\centering
\begin{tabular}{l|rrr|rrr}
\hline
& \multicolumn{3}{c|}{$D(\cG_\cS)/D_{min}(\cG)$} & \multicolumn{3}{c}{\% Error} \\
\hline
$|S|$ & 10 & 100 & 1000 & 10 & 100 & 1000 \\
\hline
LVJ & 1.0112 & 1.0110 & 1.0183 & 1.12 & 1.10 & 1.83 \\
PTN & 1.1684 & 1.0859 & 1.0790 & 16.84 & 8.59 & 7.90 \\
MCO & 1.0375 & 1.0668 & 1.0435 & 3.75 & 6.68 & 4.35 \\
CTS & 1.0526 & 1.0438 & 1.0138 & 5.26 & 4.38 & 1.38 \\
\hline
\end{tabular}
\end{table}

\section{Related Work}
\label{section:related_work}

The Steiner minimal tree problem is one of Karp's 21 NP-complete problems~\cite{Karp1972}. Hakimi~\cite{Hakimi1971SteinersPI} is often credited for proposing the first exact solution; followed by a significant number of contributions over the decades -- a comprehensive survey is available in \cite{Hauptmann13acompendium}. Often practical applications solve a particular variant of the problem, such as the Steiner arborescence, euclidean and rectilinear minimum tree, group, prize-collecting, and node-weighted Steiner tree problem~\cite{Hauptmann13acompendium, Rehfeldt2015AGA}. GeoSteiner~\cite{Warme2000} is a well known solver for exact euclidean and rectilinear Steiner minimal tree problems. SCIP-Jack~\cite{
Rehfeldt.2021.10.1007/978-3-030-55240-4_10, Rehfeldt.2021.10.1007/978-3-030-73879-2_33
} is the state-of-the-art general purpose, exact solver, winner of the 11'th DIMACS challenge, in the rooted prize-collecting 
problem category. SCIP-Jack follows a branch-and-cut approach which enables tight liner programming (LP) relaxation. 

For many years, there have been continuing interests in polynomial-time algorithms with tight approximation bound. In \S\ref{section:introduction} -- \ref{section:solution_approach}, we discussed key 2-approximation algorithms. 
Winter and Smith~\cite{Winter.1992.c29ec94074cf11dbbee902004c4f4f50} proposed heuristics with approximation bound $\leq 2$, however, they do not improve runtime over~\cite{kou.1981.10.1007/BF00288961} and are not good candidate for parallelization. More recently, Hougardy and Pr\"omel improved the approximation ratio to 1.598~\cite{Hougardy99a1.598}, followed by Robins and Zelikovsky further improving the ratio to 1.55~\cite{Robins00improvedsteiner}. T.t.b.o.o.k., to date, the LP-based approximation algorithm based on iterative randomized rounding by Byrka et al.~\cite{Byrka.2010.stoc.10.1145/1806689.1806769} offers guarantees for the best approximation ratio: $ln(4) + \varepsilon \leq 1.39$. 
Often algorithms with approximation ratio $<2$ iteratively refine a base-solution which is typically computed using a 2-approximation algorithm~\cite{Park.2008.4682068}. 
STAR~\cite{Kasneci.2009.ICDE.10.1109/ICDE.2009.64} and SketchLS~\cite{Gubichev.2012.CIKM.10.1145/2396761.2398460}, although lack guarantees of polynomial runtime and tight approximation bound, were evaluated using real-world, scale-free graphs. Genetic algorithm~\cite{Kapsalis.1993.GA.SMT}, Swarm optimization~\cite{Ma.2010.5583217}, Physarum optimization~\cite{Sun.2016.10.1109/CEC.2016.7744201} heuristics have been used for approximating Steiner minimal trees -- they only offer probabilistic solutions and their result quality suffers with increasing graph size.

We are not aware of any parallel, distributed 
solution that can compete with the scalability demonstrated in this paper. Several papers discuss parallel, distributed Steiner tree computation from the theoretical perspective only~\cite{Park.2008.4682068, Makki.1993.315344, Saikia.2020.doi:10.1142/S0129054120500367, Akbari.2007.ICPADS.4447726}. PARSTEINER94~\cite{Harris1998.HBCO.SMT}, algorithmically similar to GeoSteiner, is an early parallel exact solver. 
ParaSCIP~\cite{GamrathKochRehfeldtetal.2014} is an MPI library enabling SCIP-Jack to harness distributed clusters: it replicates the data graph on all processes and performs parallel branch-and-bound search. A distributed dual ascent~\cite{Winter.1992.c29ec94074cf11dbbee902004c4f4f50} heuristic is presented in~\cite{Drummond.2009.10.5555/1527155.1527158} whose performance was demonstrated using graphs with less than 10K edges. A GPU implementation of the STAR heuristic~\cite{Kasneci.2009.ICDE.10.1109/ICDE.2009.64} is presented in~\cite{Mathieu.10.1007/978-3-319-27122-4_31}. Group Steiner tree computation on the GPU with application to VLSI routing is presented in~\cite{Maringanti.2017.JCSE.2153-4136}. Curious readers are encouraged to check out the survey paper by Bezensek and Robic~\cite{Bezensek.2014.IJPP.DASTP} -- it presents an extensive survey of distributed Steiner tree computation efforts.

\section{Conclusion}
\label{section:conclusion}

This paper presents a parallel Steiner tree algorithm and its evaluation using a proof-of-concept distributed implementations. We demonstrate the ability to generate Steiner trees within 2-approximation of the Steiner minimal tree for thousands of seed vertices in graphs with up to 128 billion edges, t.t.b.o.o.k., to date, the largest graph scale for this problem. We show up to 90\% efficient strong scaling and present analyses that highlight the benefits of our design choices. Finally, empirical comparison with the state-of-the-art exact Steiner minimum tree solver confirms the fidelity of our solution.

\section*{Acknowledgement}
\label{section:acknowledgment}

Lawrence Livermore National Laboratory is operated by Lawrence Livermore National Security, LLC, for the U.S. Department of Energy, National Nuclear Security Administration under Contract DE-AC52-07NA27344. Funding from project LDRD \#21-ERD-020 was used in this work.

\bibliographystyle{IEEEtran}
\bibliography{ipdps_ieee_conf_main}



%




\end{document}